\documentclass[aps,preprint]{revtex4}
\usepackage[dvips]{graphics,graphicx}
\usepackage{amsmath}

\begin{document}

\title{Escape dynamics of Bose-Hubbard dimer out of a trap}
\author{Dmitrii N. Maksimov$^1$ and Andrey R. Kolovsky$^{1,2}$}
\affiliation{$^1$Kirensky Institute of Physics, 660036 Krasnoyarsk, Russia}
\affiliation{$^2$Siberian Federal University, 660041 Krasnoyarsk, Russia}
\date{\today}

\begin{abstract}
We consider a potential scattering of Bose-Hubbard dimer in 1D optical lattice. A numerical approach based on effective non-Hermitian Hamiltonian has been developed for solving the scattering problem. It allows to compute the tunneling and dissociation probabilities for arbitrary shape of the potential barrier and arbitrary kinetic energy of the dimer. The developed approach has been used to address the problem of two-particle decay out of a trap. In particular, it is shown that the presence of dissociation channels significantly decreases non-escape probability due to single-particle escape to those channels.
\end{abstract}
\maketitle

\section{Introduction}
\label{sec1}

In the recent decades we have seen a tremendous progress in experimental techniques for handling ultracold atoms and molecules in optical lattices \cite{Morsch, Bloch}. The optical lattices provide experimental set-ups which allow to confine nanoscale objects in one or two dimensions leading to revival of interest to low-dimensional quantum mechanics \cite{Buluta,Simo11}. Alongside, one of the most remarkable achievements   of the recent years is an unprecedented opportunity to manipulate just a few quantum objects \cite{Wuer09, Stibor, Zurn},  that creates a playground for few-body quantum theories. Amongst many interesting few-body phenomena that could be observed in optical  lattices we mention  fractional Bloch oscillations \cite{Khomeriki, Corrielli}, interband Klein tunnelling \cite{Longhi}, confinement  induced resonances in quasi-one-dimensional scattering \cite{Valiente}, bound states in continuum \cite{Zhang1}, etc.

In this work we consider the tunneling of interacting Bose atoms out of a specially engineered trap
 \cite{Paul,Dekel,Huepe,Schlagheck,Sierra,del_Campo,Pons,Longhi2,Witthaut,Zollner,Glick,Taniguchi,Rontani,Hunn13}. If the interaction is weak the mean-field approach remains a major theoretical tool to address decay and tunneling phenomena \cite{Paul, Dekel, Huepe, Schlagheck, Sierra}. Fewer attempts, however, were made to go beyond the mean field approximation utilizing Bose-Fermi duality \cite{del_Campo,Pons, Longhi2},  master equation approach \cite{Witthaut}, the multiconfiguration time-dependant Hartree method \cite{Zollner} or time-evolving block decimation numerical technique \cite{Glick}. A recent experiment \cite{Zurn} demonstrated an encouraging opportunity to observe tunneling behavior in a system of just a few atoms. In particular, it was reported that the tunneling rates deviate from predictions of uncorrelated single-particle approximation indicating the presence of pair correlations in the system. That observation was qualitatively explained in Ref.~\cite{Rontani} through quasiparticle wave-function approach. The limiting case of only two particles was considered in Ref.~\cite{Taniguchi}, where the authors analyzed two-particle decay with Coulomb interactions, and in Ref.~\cite{Hunn13}, where the authors introduced a spectral approach to tunneling decay of two interacting bosons in a lattice. The key idea of the latter paper was the exact diagonalization of two-particle Hamiltonian with asymmetric double-well potential, with the larger well playing the role of quasi-continuum.

In the present work we develop the above idea further by considering the true continuum (i.e., the size of the second well is assumed to be infinite). We formulate the problem in terms of effective non-Hermitian Hamiltonian. The idea of effective non-Hermitian Hamiltonians \cite{Dittes, Rotter} is mathematically equivalent to imposing open boundary conditions far from the scattering center. The formalism of effective non-Hermitian Hamiltonian has proved useful to describe scattering and tunneling phenomena in various branches of physics including quantum billiards \cite{Pichugin}, tight-binding chains \cite{Sadreev}, potential scattering \cite{Savin}, Bose-Hubard model \cite{Hiller, Graefe}, photonic crystals \cite{Bulgakov}. Quite recently the method was generalized for time-periodic potentials \cite{Sadreev2}. We adopt the method of effective non-Hermitian Hamiltonian to the problem of two-particle escape and show that it evaluates the decay law to a high accuracy.

Our model system consists of two interacting bosons in a lattice which are initially captured between an infinitely high wall and a potential barrier. It is known \cite{Scot94,Aubr96,Vali08} that two bosonic particles in a lattice can form a bound pair (dimer) that was observed in the fundamental  experiment by Winkler {et al.} \cite{Wink06} in 2006. The dimer can freely move across the lattice with a well-defined group velocity. If  the dimer hits a potential barrier or a well it can be reflected, tunnel through the barrier as the whole, or dissociate into two  independent bosons \cite{89}. In the later case, to satisfy the energy conservation, one of bosons stays in the potential well  (the case of attractively interacting bosons) or at the potential barrier (repulsive interactions). In the above cited paper   \cite{89} the tunneling and dissociation probabilities were found by simulating wave-packet dynamics of the dimer   (see also \cite{Puetter} for analogous work on fermionic systems). These numerical simulations become more and more    time consuming when the dimer kinetic energy approaches the bottom or top of the energy band, due to decrease of the   group velocity. For this reason the analysis of Ref.~\cite{89} was restricted to the middle of the energy band.  In Sec.~\ref{sec2} of the present work we formulate the problem of dimer tunneling as a stationary scattering problem.   This allows us to find the tunneling and dissociation probabilities for arbitrary quasimomentum of the incoming dimer and,   importantly, with essentially less numerical efforts than the wave-packet simulations. These results provide the basis for studying   more complicated problem of tunneling out of trap, Sec.~\ref{sec3}. We shall show that the two-particle decay is generally non-exponential and strongly dependent on details of the initial state of the dimer which one might naively consider as unimportant.

\section{ $ \mathcal{S} $-matrix theory}
\label{sec2}

To be specific, we consider a system of two attractively interacting bosons which are loaded into a 1D lattice containing a potential well. The dynamics is controlled by the Bose-Hubbard Hamiltonian
\begin{equation}
\label{BH}
\widehat{\mathcal{H}}=-\frac{J}{2} \sum_{m=-\infty}^{\infty} (\widehat{b}^{\dagger}_{m+1}\widehat{b}_{m}+
\widehat{b}^{\dagger}_{m-1}\widehat{b}_{m})
+\sum_{m=-\infty}^{\infty} v_m\widehat{n}_m +
\frac{U}{2}\sum_{m=-\infty}^{\infty}\widehat{n}_{m}(\widehat{n}_{m}-1) \;,
\end{equation}
where $\widehat{b}^{\dagger}_{m}$ and $\widehat{b}_{m}$ are standard bosonic creation and annihilation operators,
$\widehat{n}_{m}=\widehat{b}^{\dagger}_{m}\widehat{b}_{m}$ is the number operator, $J$ the hopping matrix element, $U$ the interaction constant ($U<0$),  and the on-site potential $v_m$ describes a localized well. 

\subsection{Scattering channels}

\begin{figure}[h]
\begin{center}
\includegraphics[width=0.4\textwidth]{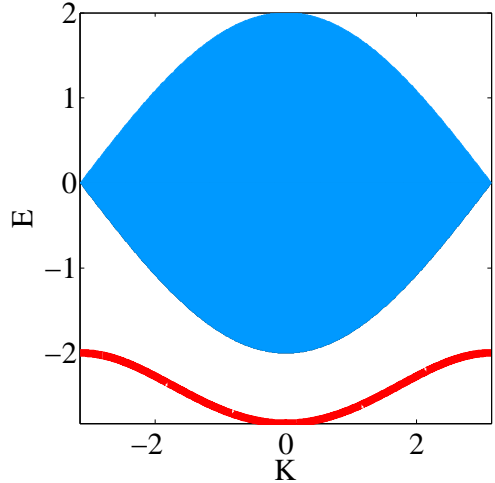}
\caption{(Color online) Dispersion of the bound pair (red line) and  scattering continuum (blue-shaded area). Parameters are $J=1$ and  $U=-2$. Through the paper we use dimensionless quantities where the energy and parameters of the Bose-Hubbard Hamiltonian are measured in units of the hopping energy (thus $J=1$) and the quasimomentum in units of the inverse lattice period.}
\label{fig0}
\end{center}
\end{figure}

We start with rewriting the eigenvalue problem for two-particle Bose-Hubbard Hamiltonian (\ref{BH}) in the form of 2D Sch\"odinger equation
\begin{equation}
\label{Hamiltonian}
-\frac{J}{2}(\Psi_{m+1,n}+\Psi_{m-1,n}+\Psi_{m,n+1}+\Psi_{m,n-1})
+(v_m+v_n)\Psi_{n,m}+ U\delta^{m}_{n} \Psi_{n,m}=E\Psi_{n,m},
\end{equation}
where $m,n$ are the coordinates of the particles. The wave function $\Psi_{n,m}\equiv\Psi(n,m)$ is symmetric with respect to permutation of the particle coordinates, i.e., $\Psi(n,m)=\Psi(m,n)$. To formulate the scattering problem we need to know asymptotic solutions of the Sch\"odinger equation (\ref{Hamiltonian}). For vanishing scattering potential the energy spectrum of the bound pair  is given by the following equation \cite{Scot94,Aubr96,Vali08}
\begin{equation}
\label{disp1}
E=-2J\cos(K/2)\sqrt{1+ {\left(\frac{U}{2J\cos(K/2)}\right)}^2}.
\end{equation}
It corresponds to a traveling wave solution of Eq.~(\ref{Hamiltonian})
\begin{equation} \label{dimer}
\Psi^{(\pm)}(m,n)=\sqrt{\frac{\sinh(\lambda)}{J\sin(K/2)}}e^{\pm iK(m+n-N)/2-\lambda|m-n|},
\end{equation}
where $\lambda$ is defined through
\begin{displaymath}
2J\sinh(\lambda)\cos(K/2)=-U  \;,
\end{displaymath}
and $K$ is quasimomentum in the center of mass reference frame. Notice that the solution Eq.~(\ref{dimer}) is normalized to a unit probability current. In what follows we set $U=-2$ and $J=1$ to ensure that the dimer propagation band does not overlap with the scattering continuum of unbound two particle solutions. The dispersion law Eq.~(\ref{disp1}) is shown in Fig.\ref{fig0} along with the shaded area of the scattering continuum.

Next we introduce dissociation channels. Let us assume that the potential $v_m\equiv v(m)$ supports a number of localized single-particle states with the energies $E_b$ below the single particle propagation band, $E_b<-J$. We require that all bound states are localized within the domain $[-N, N]$. Obviously, one can always choose $N$ large enough to fulfill the above requirement. Denoting the localized states by $\psi_b$, the wave function of the dissociation channel with one of the particles far away from the scatterer can be written as
\begin{equation}
\label{dissociation}
\Phi^{(b)}_{L,R}(m,n)=\frac{e^{\pm i k_bN}}{\sqrt{2J|\sin(k_b)|}}
\left[\psi_b(n)e^{\mp i k_b m} \Theta(\mp m - N)
+ \psi_b(m)e^{\mp i k_b n} \Theta(\mp n - N) \right],
\end{equation}
where indices $L,R$ denote the waves travelling to the left (right) from the scattering region, $\Theta(n)$ is the Heaviside function
\begin{equation}
\label{Heviside}
\Theta(n)= \left\{
\begin{array}{cc}
0 & \mbox{if $n \leq 0$}; \\ 1 & \mbox{if $n > 0$},
\end{array} \right.
\end{equation}
and the wave number $k_b$ is found from the dispersion relation
\begin{equation}
\label{disp2}
E=E_b-J\cos(k_b),
\end{equation}
where $E$ is the dimer energy (\ref{disp1}). Notice that $k_b$ found from Eq.~(\ref{disp2}) is not always real. If $k_b$ is not real the equation (\ref{dissociation}) should be interpreted as an evanescent wave which decays exponentially away from the scattering center. Thus, the number of the dissociation channels, which we label by the index $b$, varies with the energy $E$ of the scattered dimer.

\subsection{Matching asymptotic solutions}

In the presence of the scattering potential the dimer reflection and transmission channels are obviously given by Eq.~(\ref{dimer}) multiplied by the Heaviside function:
\begin{equation}
\label{bound}
\Psi_{L,R}=\Psi^{(\mp)}(m,n)[1-\Theta(N \pm n  )\Theta(N \pm m  )].
\end{equation}
Now, let us assume that the incident wave is superposition of incoming two-particle states
\begin{equation}
\label{in}
\Psi_{in}=a_L\Psi_L^{*}+ a_R\Psi_R^{*} + \sum_{b=1}^{N_b} a^{(b)}_L{(\Phi_L^{(b)})}^{*} +
\sum_{b=1}^{N_b} a^{(b)}_R{(\Phi_R^{(b)})}^{*}.
\end{equation}
Notice that the above equations contains waves incident through dissociation channels. It could be physically interpreted as a collision between a single boson with another boson already captured in the scattering center. The solution of the scattering problem can be presented in the following form
\begin{equation}
\label{solution}
\Psi= \Psi_{in}+ c_L\Psi_L+ c_R\Psi_R + \sum_{b=1}^{N_b} c^{(b)}_{L}\Phi_L^{(b)} +
\sum_{b=1}^{N_b} c^{(b)}_{R}\Phi_R^{(b)} +\sum_{p,q=-N}^{N}\chi_{p,q}\phi_{p,q},
\end{equation}
where $\phi_{p,q}=\phi_{p,q}(n,m)$ is a complete set of basis functions in the box $-N\leq n,m \leq N$. In this work we shall use the number states as the basis, i.e.,
\begin{equation}
\phi_{p,q}(n,m)=\delta_{m}^{p}\delta_{n}^{q}.
\end{equation}
Notice that within the box $-N\leq n,m \leq N$ Eq.~(\ref{solution}) includes all possible degrees of freedom whose contributions come with yet unknown coefficients $\chi_{p,q}$. Outside the box the solution is expanded over all possible scattering channels. The key idea of our approach is to use the exact representation of the Bose-Hubbard Hamiltonian within the box, where the scattering occurs, while outside the box the solution is projected onto the channel functions Eqs.~(\ref{dissociation},\ref{bound},\ref{in}).

Let us use the symbol $\phi_j$ for the $j_{\mathrm{th}}$ function from a set $(\phi_{p,q}, \Psi_{L,R}, \Phi^{(b)}_{L,R})$. To be more specific, in this set we have $(2N+1)^2$ functions $\phi_{p,q}$ accounting for dynamics within the scattering region, two dimer channel functions, and $2N_b$ functions for the dissociation channels. Since the wave function $\Psi$ satisfies the stationary Schr\"{o}dinger equation
\begin{equation}
\label{Shr}
(\widehat{\mathcal{H}}-E)\Psi=0,
\end{equation}
the evaluation of the scalar products $\langle\phi_j|(\widehat{\mathcal{H}}-E)|\Psi \rangle=0$ yields a set of $(2N+1)^2+2(N_b+1)$ linear equations for variables $\chi_{p,q}, c_L, c_R, c_L^{(b)}, c_R^{(b)}$. Assuming for a moment that the well supports only one single-particle bounds state, after some elementary but tedious algebra one finds a matrix equation in the following form
\begin{equation}
\label{matrix}
 \left\{
\begin{array}{ccccc} \widehat{\mathcal{H}}_0-E & W_L & W_R & V_L & V_R
\\ W^{\dagger}_L & P & 0 & 0 & 0
\\ W^{\dagger}_R & 0 & P & 0 & 0
\\ V^{\dagger}_L & 0 & 0 & Q_1 & 0
\\ V^{\dagger}_R & 0 & 0 & 0 & Q_1
\end{array}
\right\} \left\{
\begin{array}{c}
|\chi\rangle \\ c_L \\ c_R \\ c_L^{(1)} \\ c_R^{(1)}
\end{array}
\right\}= \left\{
\begin{array}{c}
-\Theta \\ -Ga_L \\ -Ga_R \\ -Q_1a_L^{(1)} \\ -Q_1a_R^{(1)}
\end{array}
\right\}
\end{equation}
Here $\widehat{\mathcal{H}}_0$ is the sub-block describing the couplings among the interior degrees of freedom and $|\chi\rangle$ is a vector of coefficients $\chi_{p,q}$. It is easily seen that $\widehat{\mathcal{H}}_0$ is nothing but the Bose-Hubbard Hamiltonian Eq.(\ref{BH}) in the matrix form Eq.~(\ref{Hamiltonian}). The source term $\Theta$ is given by
\begin{equation}
\Theta=\sum_{C=L,R} W^{*}_{C}a_C +\sum_{C=L,R} {V}_{1,C}^{*} a_{C}^{(1)}.
\end{equation}
The coupling between the interior degrees of freedom and the dimer reflection  (transmission) channel is accounted for by $(2N+1)^2\times1$ matrix $W_{L,R}$. The only nonzero elements of $W_{L,R}$ are given by
\begin{equation}
{(W_{L,R})}_{m,n}=-\frac{e^{iK/2}}{2}\sqrt{\frac{J\sinh(\lambda)}{\sin(K/2)}}
\left(\delta^{\mp N}_{m}e^{ iKn/2-\lambda|\mp n-(N+1)|}
+\delta^{\mp N}_{n}e^{iKm/2-\lambda|\mp m-(N+1)|}\right).
\end{equation}
The scalars $P$ and $G$ are found as
\begin{equation}
P=\frac{e^{-iK/2}}{2\sin(K/2)}, \ \
G=\frac{\sinh(\lambda)e^{-iK/2}}{\sin(K/2)}
\frac{e^{-iKN
}}{e^{\lambda}-e^{iK-\lambda}} .
\end{equation}
Coupling to the left (right) dissociation channel is described by $(2N+1)^2\times1$ matrix $V_{L,R}$. The nonzero elements of $V_{L,R}$ are
\begin{equation} \label{disch}
{(V_{L,R})}_{m,n}=\frac{-1}{2}\sqrt{\frac{J}{2\sin(k_b)}} e^{ik}
\left[ \psi_b(n)\delta^{m}_{\mp N} +\psi_b(m)\delta^{n}_{\mp N} \right]
\end{equation}
while $Q_b$ is given by
\begin{equation} \label{Q}
Q_b=\frac{e^{-ik_b}}{2\sin(k_b)}.
\end{equation}
In case the system allows many dissociation channels Eq.~(\ref{matrix}) should be complemented with additional rows and columns whose elements are evaluated according to (\ref{disch}) and (\ref{Q}) for each localized single-particle bound state available for occupation. Thus, in general $c^{(1)}_L$ and $c^{(1)}_R$ should be replaced with $N_b\times 1$ matrices composed of reflection amplitudes $c^{(b)}_L$ and $c^{(b)}_R$, while each $Q_1$ is replaced with $N_b\times N_b$ diagonal matrix with $Q_b$ on the main diagonal. The source term $\Theta$ then reads
\begin{equation}
\Theta=\sum_{C=L,R} W^{*}_{C}a_C +\sum_{b=1}^{N_b}\sum_{C=L,R} {V}_{b,C}^{*} a_{C}^{(b)}.
\end{equation}

\subsection{Effective non-Hermitian Hamiltonian}

In principle, Eq.~(\ref{matrix}) is already sufficient to find the tunneling and dissociation probabilities. Nevertheless, it is useful to formalize the problem
further, which leads to the notion of the effective non-Hermitian Hamiltonian. In the next step we eliminate variables $ c_L, c_R, c_L^{(b)}, c_R^{(b)}$ from Eq.~(\ref{matrix}) which could easily done thanks the variables $P$ and $Q_b$ being scalar quantities. First, Eq. (\ref{matrix}) is solved for $c_L, c_R, c_L^{(b)}, c_R^{(b)}$, and then the resulting expressions are substituted into the first row of Eq. (\ref{matrix}) to yield an algebraic equation for the interior wave function $|\chi\rangle$ as
\begin{equation}
\label{scattering}
(\widehat{\mathcal{H}}_{eff}-E)|\chi\rangle= \sum_{C=L,R} [f(K)W_C-W^{*}_{C}]a_C
-i\sum_{b=1}^{N_b}\sqrt{2\sin{k_b}}\sum_{C=L,R} \widetilde{V}_{b,C} a_{C}^{(b)} \;,
\end{equation}
where
\begin{equation}
{(\widetilde{V}_{L,R})}_{m,n}=-\frac{\sqrt{J}}{2}
\left[ \psi_b(n)\delta^{m}_{\mp N} +\psi_b(m)\delta^{n}_{\mp N} \right],
\end{equation}
and
\begin{equation}
f(K)=\frac{G}{P}=\frac{2\sinh(\lambda)e^{-iKN}}{e^{\lambda}-e^{iK-\lambda}} \;,
\end{equation}
while operator $\widehat{\mathcal{H}}_{eff}$ has the following form
\begin{equation}
\label{effective}
\widehat{\mathcal{H}}_{eff}=\widehat{\mathcal{H}}_0-\sinh{(\lambda)}\sum_{C=L,R} \widetilde{W}_C\widetilde{W}_C^{\dagger}
 e^{iK/2} - \sum_{b=1}^{N_b} \sum_{C=L,R} \widetilde{V}_{b,C}\widetilde{V}_{b,C}^{\dagger}e^{ik_b},
\end{equation}
with ${(\widetilde{W}_{L,R})}_{m,n}$ given by
\begin{equation}
{(\widetilde{W}_{L,R})}_{m,n}=-\sqrt{\frac{J}{2}}
\left(\delta^{\mp N}_{m}e^{ iKn/2-\lambda|\mp n-(N+1)|}
+\delta^{\mp N}_{n}e^{iKm/2-\lambda|\mp m-(N+1)|}\right) \;.
\end{equation}
The operator (\ref{effective}) could be easily  recognized as effective non-Hermitian Hamiltonian \cite{Sadreev}. It  has the structure typical for
equations describing the systems with an open boundary such as the coupled mode theory equations \cite{Fan}, although written in the coordinate rather then in the energy representation. One of the most important features is the emergence of factors $e^{iK/2}$ and $e^{ik_b}$ accounting for the band structure of the continua, which is again consistent with the single-particle tight-binding theory \cite{Sadreev}.

It is instructive to rewrite the effective non-Hermitian Hamiltonian in terms of creation and annihilation operators. We have
\begin{equation}
\widehat{\mathcal{H}}_{eff}=\widehat{\mathcal{H}}_{0} - \sum_\pm  \left[\sum_{m'=-N}^N
\left( e^{iK/2}\zeta_{\kappa}(m',m) \widehat{b}_{m'}^{\dagger}\widehat{b}_{m}
+\frac{J}{2}\sum^{N_b}_{b=1}e^{ik_b} \psi_{b}(m') \psi_{b}(m) \widehat{b}_{m'}^{\dagger}\widehat{b}_{m}
\right)\right] \widehat{n}_{\pm N} ,
\end{equation}
where
\begin{equation}\label{Heff2}
\zeta_{\mp N}(m',m)=\frac{J(2-\delta_{m'}^{m})}{2}\sinh(\lambda)e^{ iK(m-m')/2-\lambda|\mp m-(N+1)|-\lambda|\mp m'-(N+1)|}
\end{equation}
We would like to point out that unlike in the previous studies \cite{Hiller, Graefe}, where the effective non-Hermitian Hamiltonian was introduced
phenomenologically by including the decay term $i\widehat{n}_{\pm N}$, here we obtain it from the first principles. One can see that in the full-fledged formulation the anti-Hermitian term is non-local albeit in the case of the dimer scattering channel it decays exponentially away from the truncation site $N$. Furthermore, the non-Hermitian Hamiltonian is proved to be dependent on the spectral parameters of the scattering channels. We would like to stress that the resulting expression for the effective non-Hermitian Hamiltonian is exact. Formally it corresponds to reflectionless boundary conditions. This allows to avoid spurious reflection which are typical for complex absorbing potentials,  that is known to distort the decay dynamics \cite{Muga}. One the other hand, the fact that the exact reflection-less potential could be both energy-dependent and non-local is in compliance with findings on atom detection by fluorescence \cite{Ruschhaupt}.

\subsection{$\mathcal{S}$-matrix}

Using the solution of Eq.~(\ref{scattering}) for the interior wave function we can find explicit expression for the scattering matrix. The $\mathcal{S}$-matrix is defined through an equation connecting the vectors of incoming $A^{T}=(a_L,a_R,a^{(b)}_L,a^{(b)}_R)$ and outgoing  amplitudes $B^{T}=(c_L,c_R,c^{(b)}_L,c^{(b)}_R)$,
\begin{equation}
B={\mathcal{S}}A.
\end{equation}
Let us denote by  $|\chi_{\tau}\rangle$ the interior solution produced via population of a single incoming channel $\tau$. Then for the reflection into dimer channels (i.e $\tau'=1,2$) Eq.(\ref{matrix}) yields
\begin{equation}
\mathcal{S}_{\tau',\tau}=\Delta_{\tau',\tau}-\sqrt{2\sinh(\lambda)\sin(K/2)}\widetilde{W}_{\tau'}^{\dagger}|\chi_{\tau}\rangle,
\end{equation}
while for reflection into dissociation channels ($\tau'>2$)
\begin{equation}
\mathcal{S}_{\tau',\tau}=\Delta_{\tau',\tau}-\sqrt{2\sin{k_{b_{\tau'}}}}\widetilde{V}_{\tau'}^{\dagger}|\chi_{\tau}\rangle,
\end{equation}
where $\Delta_{\tau',\tau}$ is a diagonal matrix
\begin{equation}
\Delta_{\tau',\tau}={\rm diag}[-f(K),-f(K),-1,-1,\dots,-1,-1].
\end{equation}

\subsection{Numerical example}

\begin{figure}[t]
\begin{center}
\includegraphics[width=0.4\textwidth]{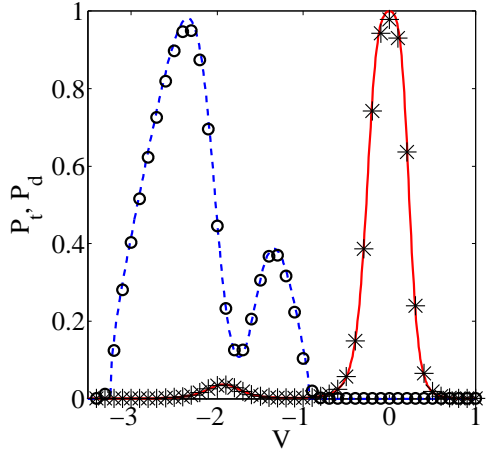}
\caption{(Color online) Co-tunneling $P_t$ (red solid line) and dissociation  $P_d$ (blue dashed line) probabilities for the bound pair as the functions of the
parameter $V$ in Eq.~(\ref{potential}). The quasimomentum of the incident pair $K=\pi/2$. Circles and stars show results of  Ref.~\cite{89}.}
\label{fig1}
\end{center}
\end{figure}

\begin{figure}[t]
\begin{center}
\includegraphics[width=0.45\textwidth]{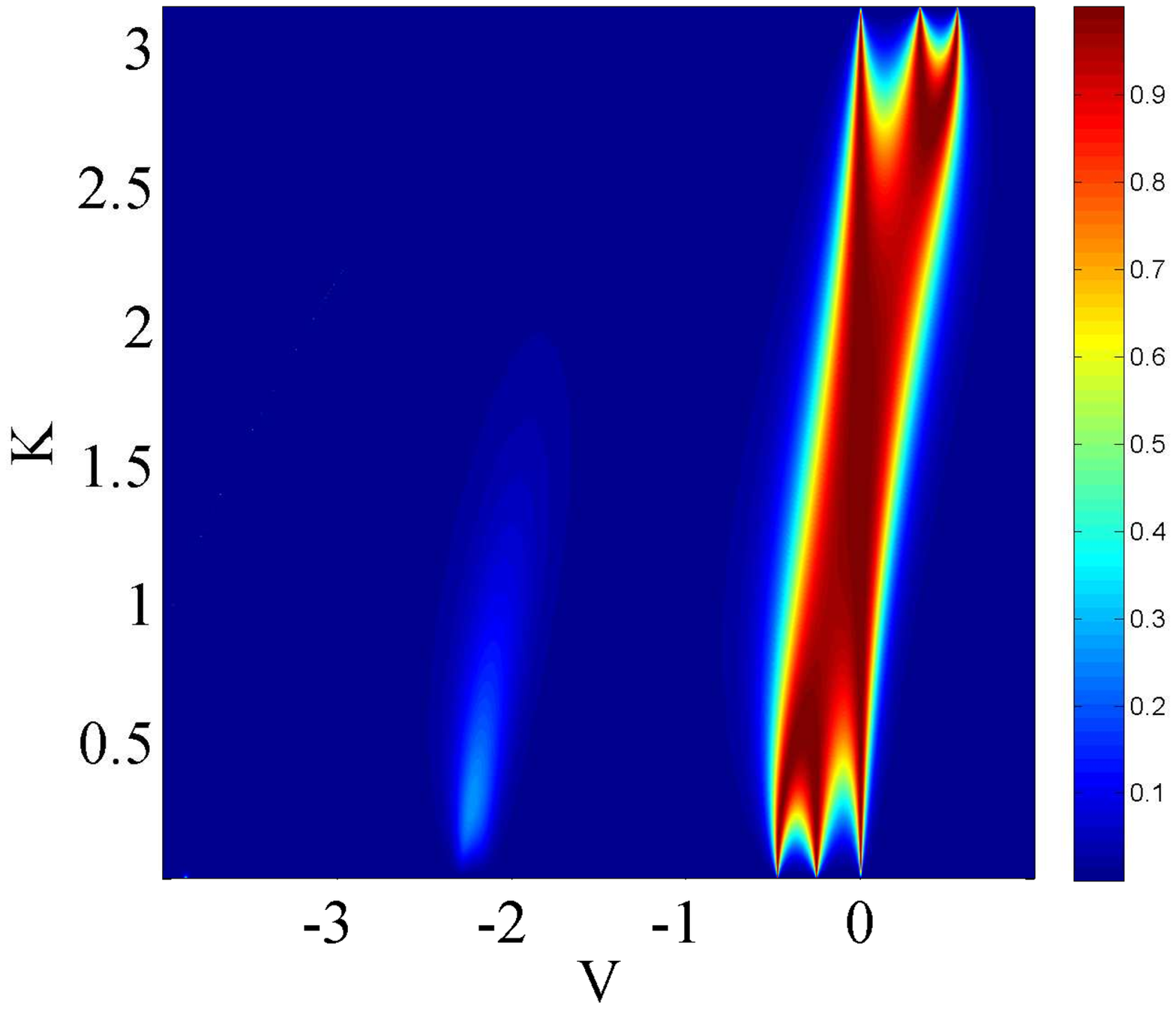}
\includegraphics[width=0.45\textwidth]{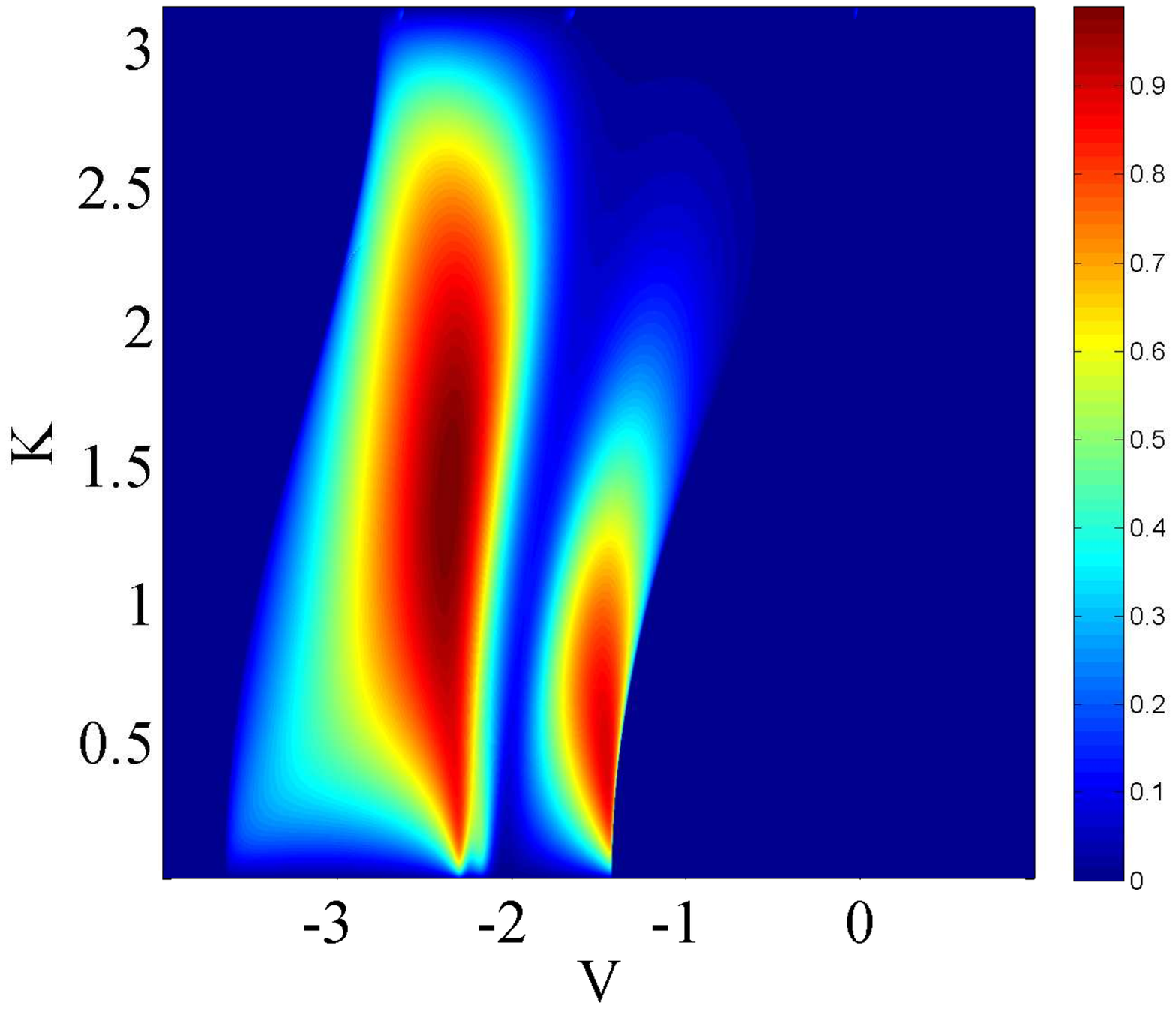}
\caption{(Color online) Co-tunneling $P_t$ (left) and dissociation $P_d$ (right) probabilities as the function of barrier height $V$ and the dimer quasimomentum $K$.}
\label{fig2}
\end{center}
\end{figure}

To test our method we solved the scattering problem with potential $v(m)$ given by
\begin{equation}
\label{potential}
v(m)=Ve^{-m^2/2\sigma^2},
\end{equation}
with $\sigma=0.65$. We found that for a good accuracy it is sufficient to set $N=10$. The plot of scattering probabilities vs. barrier height $V$ is shown in Fig.~\ref{fig1} for $K=\pi/2$. The depicted tunneling and dissociation probabilities fairly reproduce those obtained in Ref.~\cite{89} by using the wave-packet simulation, while the computational time decreases by two orders of magnitudes. This allows us to scan over both the quasimomentum $K$ and the height of the potential barrier $V$. The transmission $P_t$ and dissociation $P_d$ probabilities as functions of $K$ and $V$ are presented in Fig.~\ref{fig2} as a color map. These numerical results indicate that in the presence of open dissociation channels $-3<V<-1$ the bound pair tends to split with one of the particles being captured in the well rather than reflect or transmit as whole.

\begin{figure}
\begin{center}
\includegraphics[width=0.5\textwidth]{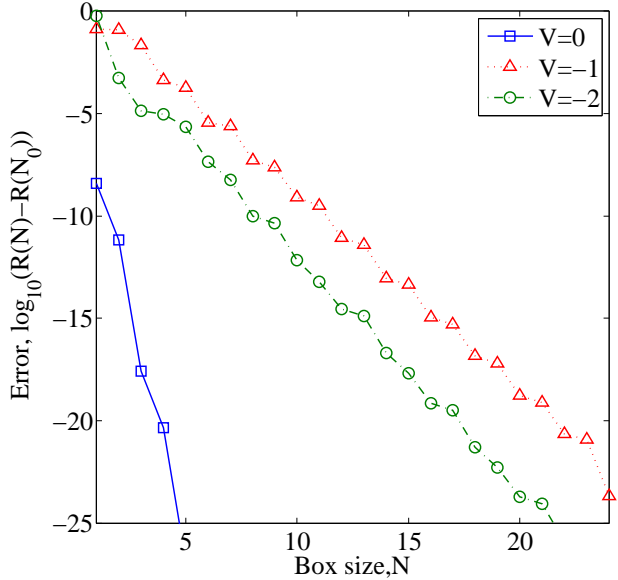}
\caption{(Color online) Logarithmic plot of the absolute error $|R(N)-R(N_0)|$ as function of truncation radius $N$ for various values of barrier height $V$. One can see that the error drops exponentially.}
\label{fig-error}
\end{center}
\end{figure}

To conclude this section we would like to make some remarks on the accuracy of the method. It is necessary in our approach that the propagation band of the dimer lies below the scattering continuum, see Fig.~\ref{fig0}. Otherwise, one would have a continuum of scattering channels and consequently formula (\ref{matrix}) would be integral rather than an algebraic equation. In our case two-particle scattering states do not propagate at energies below $-2J$. That means the contribution of the scattering continuum to the solution of Eq.~(\ref{Hamiltonian}) comes in form of evanescent waves that decay exponentially away from the scattering domain. Thus, one can conclude that as the truncation radius $N$ is increased the error would drop exponentially. To prove this we solved Eq.~(\ref{scattering}) for various values of truncation radius $N$ and evaluated corresponding reflection coefficients $R(N)$. Using $N_0=25$ as the reference point we found the error as $|R(N)-R(N_0)|$. The results are plotted in Fig.~\ref{fig-error} for three different values of barrier height $V$.

\section{Decay rates}
\label{sec3}

\begin{figure}[t]
\begin{center}
\includegraphics[width=0.5\textwidth]{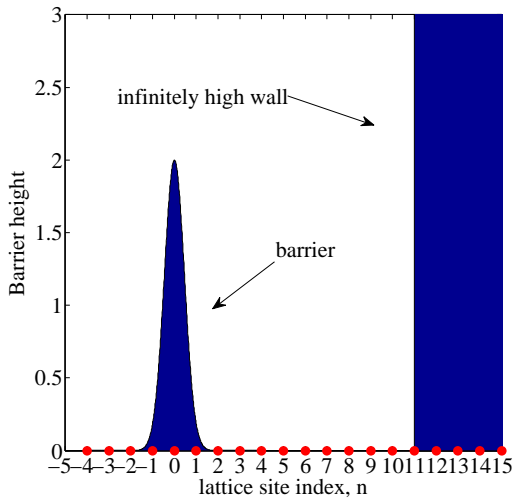}
\caption{(Color online) Configuration of a trap comprised of an infinitely high wall and a potential barrier of height $V$.}
\label{fig4}
\end{center}
\end{figure}

In this section we address the particle decay out of a trap. The trap is introduced as a length of 1D lattice confined between an infinitely high wall and a potential barrier/well of height/depth $V$, see Fig.~\ref{fig4}. In what follows we assume that the initial state of a dimer in the trap is given in the form of a Gaussian wave packet,
\begin{equation}
\label{initial}
\Psi_0=\cos(K(m+n)/2-M)e^{-{((m+n)/2-M)}^2/2-\lambda|m-n|} \;,
\end{equation}
where the parameter $M$ fixes the initial position of the dimer. The state (\ref{initial})  is well suited for the wave-packet simulations discussed later on in Sec.~\ref{sec3c}.

\subsection{Gamov's states}

The standard procedure to find the decay of a given initial states consist of two step. First one find eigenstates $\Psi_l$ and eigenvalues $z_l$ of the effective non-Hermitian Hamiltonian (\ref{effective}),
\begin{equation}
\label{eigenvalue}
\widehat{\mathcal{H}}_{eff}\Psi_l = z_l \Psi_l \;.
\end{equation}
Once the eigenstates and eigenvalues are found the initial condition (\ref{initial}) is expanded over Gamov's states $\Psi_l$,
\begin{equation}
\Psi_0=\sum_{l} B_l \Psi_l  \;.
\end{equation}
Then the imaginary part of the eigenvalue
\begin{equation}
z_l=E_l-i\frac{\gamma_l}{2}
\end{equation}
would give the lifetime of the corresponding Gamov state and the non-escape probability $\rho(t)$ would be simply given as
\begin{equation}
\label{decay}
\rho(t)=\sum_{l}{|B_l|}^2e^{-\gamma_lt}.
\end{equation}

Unfortunately, realization of this standard procedure encounters two difficulties. The first difficulty comes from the fact that, as it was shown in Ref.~\cite{Savin}, not all eigenvalues found from Eq.(\ref{eigenvalue}) correspond to the true poles of the $\mathcal{S}$-matrix. This could be understood as a consequence of the freedom in choosing the truncation radius $N$. Varying $N$ one changes the number of eigenvalues $z_l$. Nevertheless the $S$-matrix at large $N$ is asymptotically stable. This means that {\it some} of the eigenvalues $z_l$ do not correspond to the true resonances. Such emergence of spurious eigenvalues, in fact, was found to be typical for eigenvalue problems with open boundary conditions \cite{Kim}.

The second difficulty arises from the algebraic structure of $\widehat\mathcal{H}_{eff}$. As it was pointed out in the previous section $\widehat\mathcal{H}_{eff}$ itself depends on $z_l$. Hence Eq.~(\ref{eigenvalue}) could not be viewed as a standard eigenvalue problem and, in search of eigenvalues of $\widehat\mathcal{H}_{eff}$, one has to scan over the complex energy plane to minimize the norm of $\widehat{\mathcal{H}}_{eff}(z)-zI$ where $I$ is identity matrix. 

\subsection{Harmonic inversion method}

\begin{figure}[t]
\begin{center}
\includegraphics[width=0.5\textwidth]{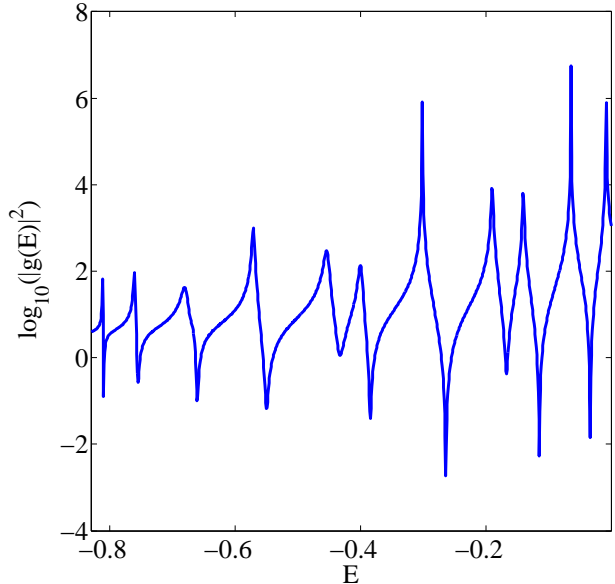}
\caption{(Color online) Logarithmic plot of the response function $G(E)$ at $V=-2$. One can see well pronounced resonant features.}
\label{fig5}
\end{center}
\end{figure}

To overcome the above difficulties we will apply the {\em harmonic inversion method} which is an efficient tool for extracting resonance positions and lifetimes from the spectral data \cite{Wall}. The method is nicely outlined in Ref.~\cite{Wiersig}. The central idea is that the response $g(E)$ of open system to an external driving is presented as the  sum
\begin{equation}
\label{inv_harm}
g(E)=\sum_{l=1}\frac{A_l}{E-\tilde{z}_l} \;,
\end{equation}
where $\tilde{z}_l$ are the complex energies corresponding to the true resonances in the system (complex poles of the scattering matrix). In this work we choose $g(E)=\tilde{\Psi}_{N_0,N_0}$, where $\tilde{\Psi}_{m,n}$ is the solution of
\begin{equation}
(\widehat{\mathcal{H}}_{eff}-E)\tilde{\Psi}_{m,n}=\delta_{N_0}^n\delta_{N_0}^m.
\end{equation}
Notice that `the external driving force' $\delta_{N_0}^n\delta_{N_0}^m$ preserves bosonic symmetry of the problem.  In our computations we took $N_0=10$. A typical dependance of the response function $g(E)$ is plotted in Fig.~\ref{fig5}. Using Eq.~(\ref{inv_harm}) we extract the resonance energies $z_l$. Finally, when the true resonances are found, we obtain the wave functions of the Gamov states solving homogenous equation (\ref{eigenvalue}).

\begin{figure}[t]
\begin{center}
\includegraphics[width=0.45\textwidth]{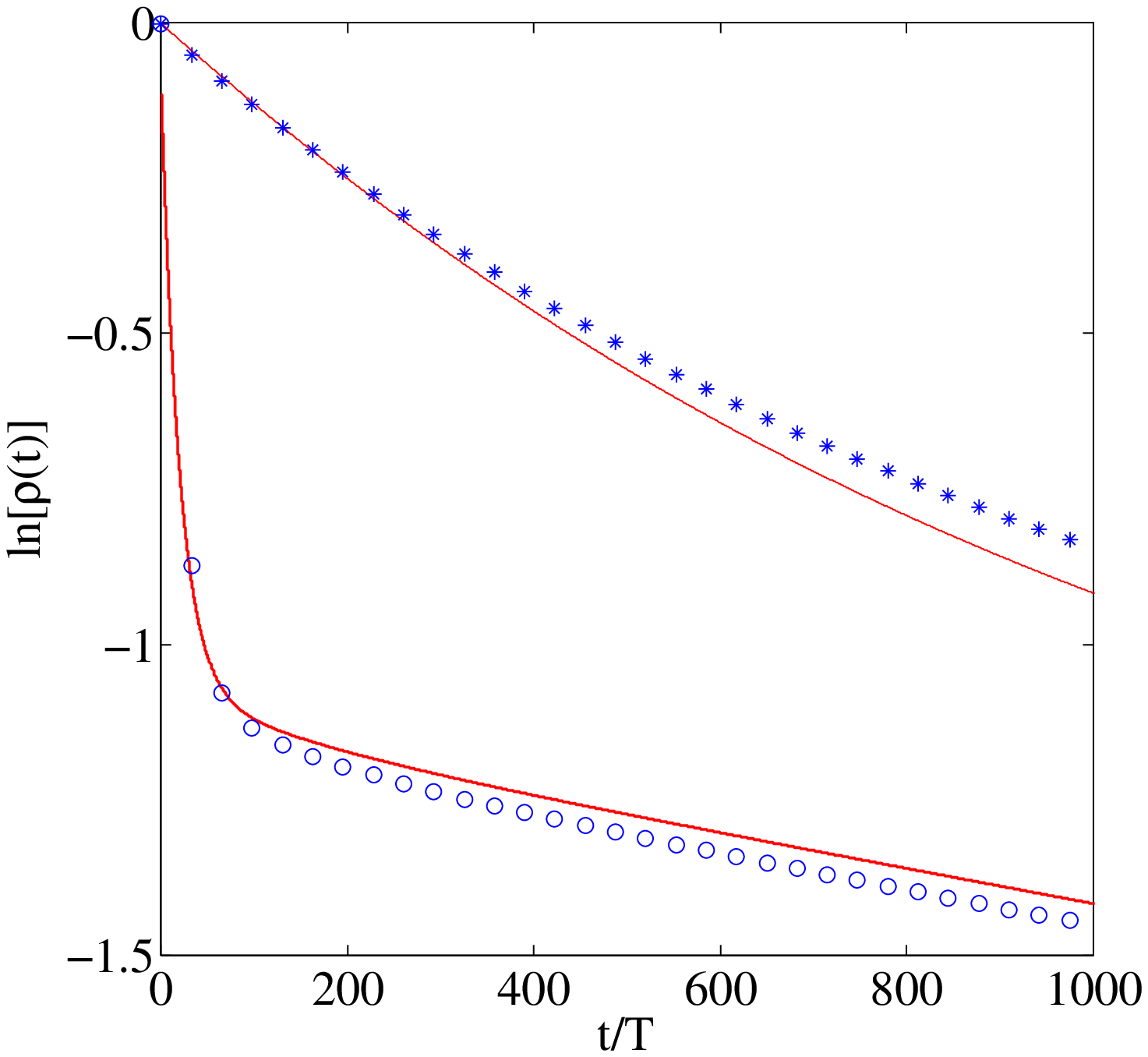}
\includegraphics[width=0.45\textwidth]{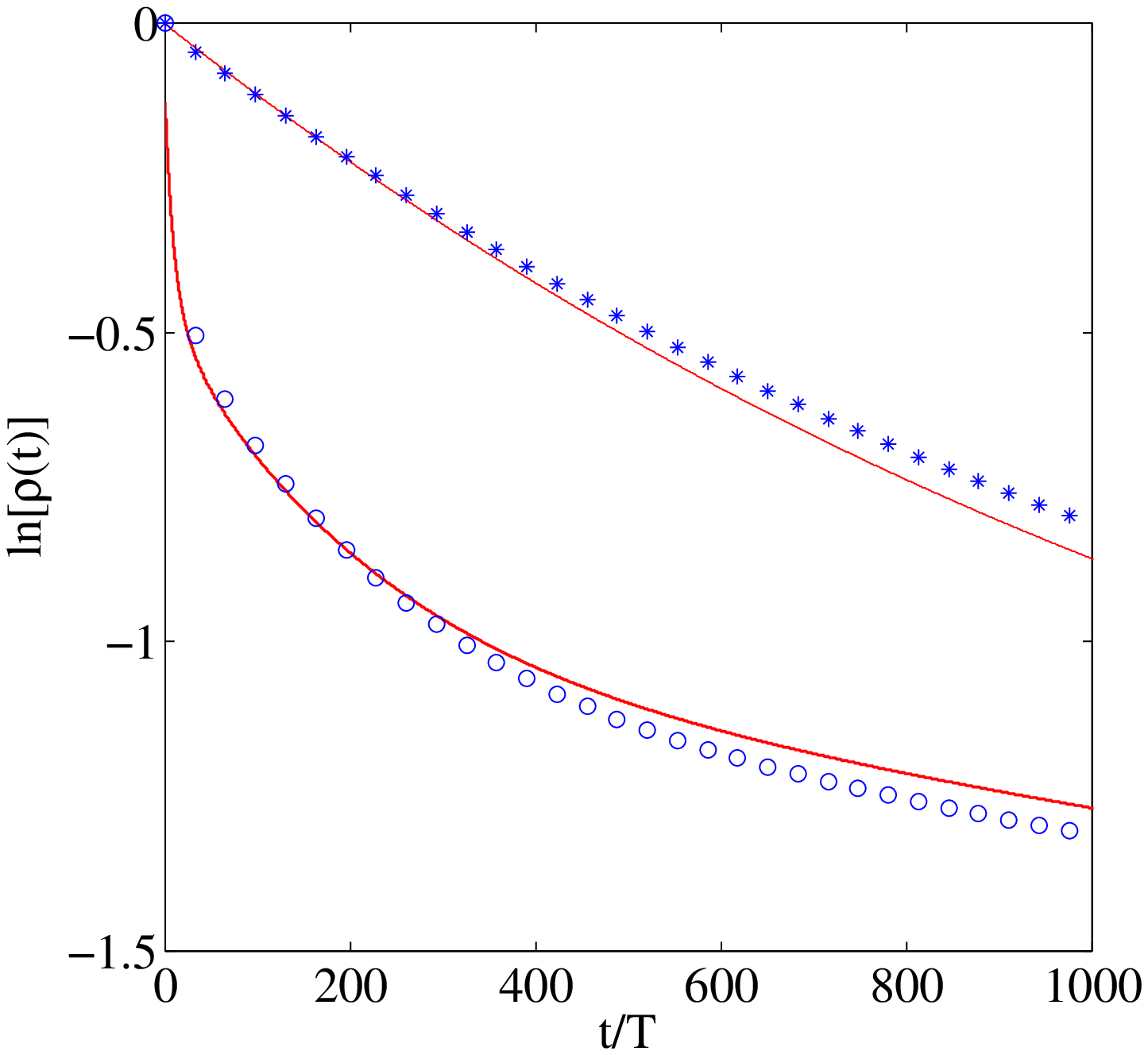}
\caption{(Color online) Non-escape probability vs. time for the trap shown in Fig.~5. The hight/depth of the potential barrier/well are $V=-2$ (open circles) and $V=0.8$ (asterisks). The initial state of the dimer is chosen in the form (\ref{initial}) with $M=5$ (left panel) and $M=6$ (right panel). Solid red lines show estimations based on Eq.~(\ref{decay}). The time is measured in units of $T=2\pi/|E|$ where  $E=-0.30$.}
\label{fig6}
\end{center}
\end{figure}

\subsection{Non-escape probability}\label{sec3c}

In this subsection we compare the result (\ref{decay}), which involves the notion of effective non-Hermitian  Hamiltonian $\widehat{\mathcal{H}}_{eff}$, with the direct numerical simulations of the escape processes, which are done by using the original Bose-Hubbard Hamilton  (\ref{BH}) with the barrier (\ref{potential}) . The corresponding time-dependent Schr\"odinger equation was solved with Crank-Nicolson method. The computational domain was truncated with the use of adiabatic absorbers \cite{Oskooi}. The results are plotted in Fig.~\ref{fig6} by symbols, where the asterisks and open circles refers to $V=-2$ and $V=0.8$, respectively.  One can see that Eq.~(\ref{decay}) reproduces the results of the direct simulations to a good accuracy.  One can also see that the presence of a dissociation channel at $V=-2$, when the dimer propagation band fully overlaps with the propagation band of the channel, drastically decreases the non-escape probability. This is consistent with the findings of the Sec.~\ref{sec2} where it was observed that the dimer tends to spit whenever a dissociation channel is accessible. In fact, the simulations show that approximately $80\%$ of the decay rate is due to the dissociation channel. In contrast at $V=0.8$ the dimer decays much slower in spite of the fact the confinement potential is weaker. We do not present results for $V=2$ because at this value of barrier height the escape probability is vanishing ($<10^{-5}$ at $t=1000T$).

The two panels in Fig.~\ref{fig6} are aimed to illustrate sensitivity of the result to seemingly unimportant parameters like, for example, the parameter $M$ which controls the initial position of the wave packet according to Eq.~(\ref{initial}). The observed, surprisingly high sensitivity to initial conditions poses the question about typical initial state or ensemble averaging. In fact, the laboratory setup for measuring non-escape probability could be as follows. Using three mutually perpendicular standing laser waves of different intensities one creates an ensemble of 1D lattices. Next, adding two sheet-like beams one creates a trap and then empty all lattice sites outside the trap by using, for example, the electron beam technique \cite{Wuer09}. If density of dimers is low enough one can also satisfy the condition that every 1D trap contains no more than one dimer. However, the initial states of these dimers are unknown and may vary from one to another 1D lattice. Thus only averaged decay rate can be measured in the laboratory experiment. We reserve the problem of the relevant ensemble of initial conditions and averaged decay dynamics for future studies.

\section{Summary and conclusion}

We considered the tunneling of a Bose-Hubbard pair of two interacting bosons through a potential barrier -- the problem addressed earlier in Ref.~\cite{89}. The results of the paper are three-fold.

First, we reformulated the problem as a stationary scattering problem for the Bose-Hubbard dimer. We developed a method which could be applied for an arbitrary asymptotically vanishing scattering potential at any value of the dimer quasimomentum. This, in particular, allows to find the conditions under which the dimer transmits, reflects, or dissociates in the process of collision with potential barrier.
It was found that the presence of dissociation channels leads to a high probability of the dimer being split with one particle captured in the scattering centre while the other is typically reflected.

Second, we derived the non-Hermitian Hamiltonian that governs the system dynamics, with the only limiting assumption that the dimer propagation band does not overlap with the scattering continuum. Unlike in the previous studies \cite{Hiller, Graefe}, where the effective non-Hermitian Hamiltonian was introduced phenomenologically by including the decay term $i\widehat{b}_N^{\dagger}\widehat{b}_N$, here we obtain it from the first principles. One can see that in the full-fledged formulation the anti-Hermitian term is non-local (albeit in the case of the dimer scattering channel it decays exponentially away from the truncation site $N$). Moreover, the full-fledged formulation comes at the price of the non-Hermitian Hamiltonian dependent on the spectral parameters of the scattering channels.

Finally, we used the developed formalism to address the problem of two-particle decay out of a trap. We proposed a recipe for finding two-particle Gamov states which give us a key to evaluating the non-escape probability. It was shown that the presence of dissociation channels substantially increases the decay rates favoring dissociation scenario, where one particle is captured in a single-particle bound state while the other leaks to the continuum. This complex tunneling process generally leads to non-exponential decay of survival probability.

Concluding, we believe that our results are relevant due to the recent progress in physics that allows creating experimental set-ups where both potential profile \cite{Meyrath, Henderson, van_Es} and interaction strength \cite{Haller, Chin} could be varied at will, and thus, could open new opportunities for engendering quantum systems with desired tunneling escape properties.

\section{Acknowledgements}

The authors acknowledge financial support of Russian Academy of Sciences through the SB RAS integration project  No.29 {\em Dynamics of atomic Bose-Einstein condensates in optical lattices}.


\end{document}